\documentclass[a4paper]{ifacconf}

\usepackage{graphicx,url}
\usepackage{svg}
\usepackage{amsmath, amsfonts, amssymb}
\usepackage{booktabs}
\usepackage[round]{natbib}
\usepackage{pgfplots}
\pgfplotsset{compat=1.18}
\def\portugues{0} 
\ifx\portugues\undefined
\def\portugues{0}
\fi

\if\portugues0
   \usepackage[english]{babel}
  \else
   \usepackage[spanish,brazil,english]{babel}
\fi

\usepackage[T1]{fontenc}

\usepackage[utf8]{inputenc}

\usepackage{ae}

\if\portugues1
 { }
 { }
 { }
 
\fi

\begin{document}

\if\portugues0

% =====================================================================
% =====================================================================
% USE THIS PART IF THE TEXT IS IN PORTUGUES OR SPANISH
% =====================================================================
% If the manuscript is in Spanish, please change the texts adequately.
% =====================================================================
% 
\selectlanguage{english}
	
\begin{frontmatter}

\title{Multi-Variable Stellar Parameter Estimation Using Residual Multitask Neural Networks \thanksref{footnoteinfo}} 

\thanks[footnoteinfo]{We thank the University of São Paulo (USP) for the financial
support provided through the Programa Unificado de Bolsas (PUB),
grant no. 2025-5561 and CNPq grant no. 404081/2023-1.}

\author{Bruno Santos Meneses Barreto, Marcio Eisencraft}

\address{Escola Politécnica, Universidade de São Paulo, SP\\
(e-mail: \{bruno.smbarreto, marcioft\}@usp.br)}

\selectlanguage{english}
\renewcommand{\abstractname}{{\bf Abstract:~}}
\begin{abstract}                % Abstract of not more than 250 words.
We present an end-to-end pipeline for estimating stellar parameters from Sloan Digital Sky Survey Data Release 12 spectra using a fully connected multitask neural network with residual blocks, whose hyperparameters are tuned via Bayesian optimization. The preprocessing pipeline includes per-spectrum standardization, \textit{RobustScaler} normalization of the target variables---effective temperature $T_{\mathrm{eff}}$, metallicity $[\mathrm{Fe/H}]$, and surface gravity $\log g$---and data augmentation via Gaussian noise injection. On a held-out test set, the model achieved Mean Absolute Errors (MAE) of 59.76~K for $T_{\mathrm{eff}}$, 0.103~dex for $[\mathrm{Fe/H}]$, and 0.130~dex for $\log g$. Normalized against the full-scale range of each parameter, these results represent range-normalized errors between 1\% and 3\%, achieved with a highly efficient model complexity of approximately 540,000 trainable parameters. These results demonstrate that a compact residual multitask architecture, combined with principled signal preprocessing, provides a parameter-efficient solution for nonlinear parameter estimation in large-scale spectral datasets. In particular, the proposed model achieves competitive performance with substantially lower complexity than deeper neural network baselines.

\vskip 1mm% não altere esse espaçamento

\end{abstract}

\selectlanguage{english}


\selectlanguage{english}

\end{frontmatter}
\else
% ===============================================================
% ===============================================================
% USE THIS PART IF THE TEXT IS IN ENGLISH
% ===============================================================
% ===============================================================
% 

\begin{frontmatter}

\title{Style for SBA Conferences \& Symposia: Use Title Case for
  Paper Title\thanksref{footnoteinfo}} 

\thanks[footnoteinfo]{Sponsor and financial support acknowledgment
goes here. Paper titles should be written in uppercase and lowercase
letters, not all uppercase.}

\author[First]{First A. Author} 
\author[Second]{Second B. Author, Jr.} 
\author[Third]{Third C. Author}

\address[First]{Faculdade de Engenharia Elétrica, Universidade do Triângulo, MG, (e-mail: autor1@faceg@univt.br).}
\address[Second]{Faculdade de Engenharia de Controle \& Automação, Universidade do Futuro, RJ (e-mail: autor2@feca.unifutu.rj)}
\address[Third]{Electrical Engineering Department, 
   Seoul National University, Seoul, Korea, (e-mail: author3@snu.ac.kr)}
   
\renewcommand{\abstractname}{{\bf Abstract:~}}   
   
\begin{abstract}                % Abstract of not more than 250 words.
These instructions give you guidelines for preparing papers for IFAC
technical meetings. Please use this document as a template to prepare
your manuscript. For submission guidelines, follow instructions on
paper submission system as well as the event website.
\end{abstract}


\end{frontmatter}
\fi

%===============================================================================
%===============================================================================
%===============================================================================

\section{Introduction}

Stellar atmospheric parameters play a central role in Astrophysics, as they provide
fundamental information about the physical properties, evolutionary state, and chemical
composition of stars, and are essential for large-scale 
studies of Galactic structure and evolution \citep{Huang_2024}. With the advent of large spectroscopic surveys such as Sloan Digital Sky Survey (SDSS) and the Large Sky Area Multi-Object Fiber Spectroscopic Telescope 
survey (LAMOST) \citep{York_2000,lamost}, which produce vast amounts of spectroscopic data, machine learning (ML) techniques have become powerful
tools for estimating these parameters accurately and at scale \citep{ivezic2014statistics}. In this work,
we investigate the use of an ML model to estimate stellar atmospheric
parameters directly from observed spectra.

Following standard stellar astrophysics definitions, the effective temperature $T_{\mathrm{eff}}$ is defined as the temperature of a blackbody that emits the same total radiative flux as the star. The metallicity $[\text{Fe/H}]$ is defined as the logarithmic ratio of the iron abundance to hydrogen abundance relative to the Sun, such that $[\text{Fe/H}] = 0$ for the Sun. The surface gravity $\log g$ represents the base-10 logarithm of the gravitational acceleration at the stellar surface (in $\mathrm{cm/s^2}$), where the solar value is approximately $\log g_{\odot} \approx 4.44$ \citep{Gray_2005}.

Stellar spectra consist of a continuum and absorption lines, and each of these atmospheric parameters affects the observed spectrum in a distinct manner \citep{Gray_2005}. In particular, $T_{\mathrm{eff}}$ governs the overall continuum shape through the spectral energy distribution, $[\mathrm{Fe/H}]$ controls the strength of absorption features via elemental abundances, and $\log g$ influences pressure broadening of spectral lines. These physically grounded relationships motivate regression methods that infer atmospheric parameters directly from spectroscopic data.

Early data-driven approaches for SDSS spectra relied on linear models, e.g., Partial Least Squares (PLS) and Least Absolute Shrinkage and Selection Operator (LASSO) \citep{10.1111/j.2517-6161.1996.tb02080.x}, which demonstrated that much of the predictive signal can be captured by low-dimensional projections of the flux vector $\mathbf{X}$ onto carefully selected wavelength regions \citep{Zhang_2009, Li_2015}. More recent methods use nonlinear function classes, e.g., The Cannon \citep{Ness_2015}, Deep Feedforward Networks \citep{2017RAA1736L}, and Convolutional Neural Networks (CNNs) \citep{10.1093/mnras/stx3298}. These approaches generally achieve higher predictive accuracy and better calibration than linear baselines. While effective, deep CNNs often impose high computational costs and require complex hyperparameter tuning.

Despite these advances, two practical challenges remain: (i) designing models that are both computationally efficient and competitive, while still being easy to train and deploy on full-resolution spectra; and (ii) evaluating them under realistic conditions rather than only on curated, high-quality subsets. This limitation is particularly evident in studies such as \cite{10.1093/mnras/stx3298}, where training was performed on a relatively small and restricted portion of the $T_{\mathrm{eff}}$ range.

In this paper, we address these challenges using a compact residual multitask multilayer perceptron (MLP) for parameter estimation from SDSS DR 12 spectra. The proposed model achieves competitive predictive performance while substantially reducing model complexity relative to deeper neural baselines.

The remainder of the paper is organized as follows. Section \ref{sec:data} describes the dataset, Section \ref{sec:preproc} presents the preprocessing pipeline, Section \ref{sec:model} details the model architecture, Section \ref{sec:train} outlines the training procedure, Section \ref{sec:Results} reports the results, and Section~\ref{sec:Conclusion} concludes the paper.
\section{DATA ACQUISITION}
\label{sec:data}

The dataset was built from the Sloan Digital Sky Survey (SDSS) Data Release 12 (DR12) catalogue \citep{Alam2015}. For each selected entry, we retrieved the corresponding FITS file URL \citep{Wells1981FITS} together with the target stellar atmospheric parameters provided by the SEGUE Stellar Parameter Pipeline (SSPP) \citep{Lee2008}. Specifically, the adopted labels were taken from the \texttt{TEFF\_ADOP}, \texttt{FEH\_ADOP}, and \texttt{LOGG\_ADOP} columns. In addition, the radial velocity information was also retrieved for use in the preprocessing stage.

The experiments were conducted on a dataset of 50,000 spectra, which was randomly partitioned into three disjoint subsets: 30,000 spectra for training, 5,000 for validation, and 15,000 for test. The split was performed after shuffling the complete dataset, so that all three partitions were drawn from the same parent distribution.

\section{Data preprocessing}
\label{sec:preproc}

Raw SDSS DR12 spectra were resampled onto a common wavelength grid for consistency across spectra. Each FITS file stores flux as a function of $\log_{10}$-wavelength (\texttt{loglam}) with approximately uniform spacing. We determined the global minimum and maximum \texttt{loglam} values across the acquired samples and defined a fixed grid with 4000 uniformly spaced points. This common grid was defined from 3.5754 to 3.9670 (3 762 \r{A} -- 9 268 \r{A}).

Each spectrum was interpolated to this common grid using \emph{cubic spline} interpolation, as implemented in SciPy \citep{scipy2020}. To avoid boundary artifacts, we applied \emph{clamping} at the ends: values outside the original wavelength coverage are set to the nearest available flux, thus avoiding undefined regions after resampling.

Table \ref{tab:stellar_params} summarizes the empirical ranges of the target parameters in the three disjoint partitions used in this study: training, validation, and test. The reported values were obtained directly from the catalog labels. The effective temperature $T_{\mathrm{eff}}$ is expressed in Kelvin, whereas $[\mathrm{Fe}/\mathrm{H}]$ and $\log g$ are expressed in dex.

\begin{table}[ht!]
\centering
\caption{Observed ranges of the target parameters in the training, validation, and test sets.}
\label{tab:stellar_params}
\begin{tabular}{lccc}
\toprule
\textbf{Split} & \textbf{$T_{\mathrm{eff}}$ (K)} & \textbf{$[\mathrm{Fe/H}]$ (dex)} & \textbf{$\log g$ (dex)} \\
\hline
Train min & 4003.46 & -4.38 & 0.18 \\
Train max & 9186.64 & 0.74 & 4.93 \\
Val min   & 4002.71 & -4.38 & 0.66 \\
Val max   & 9075.71 & 0.53 & 4.90 \\
Test min  & 4006.94 & -4.44 & 0.18 \\
Test max  & 9448.20 & 0.53 & 4.99 \\
\bottomrule
\end{tabular}
\end{table}

\begin{figure}[ht]
    \centering
    \includegraphics[width=0.33\textwidth]{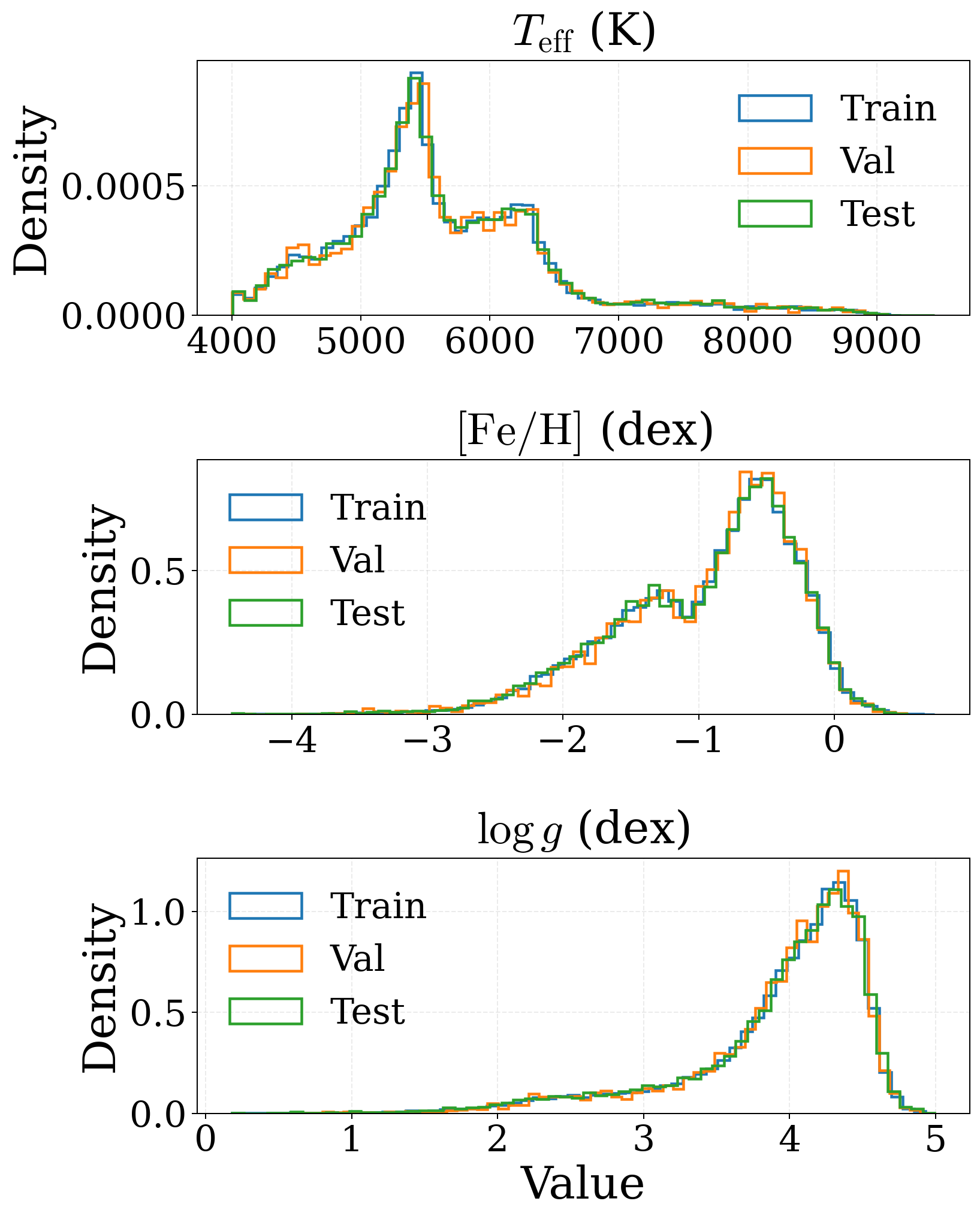}
    \caption{Density distributions of $T_{\text{eff}}$, $[\text{Fe}/\text{H}]$, and $\log g$ for the training, validation, and test sets.}
    \label{fig:dist}
\end{figure}

Figure \ref{fig:dist} illustrates the density distributions of the three target parameters across the training, validation, and test sets, indicating that the random split preserved similar distributions across the three partitions.

After resampling the spectra to a common, uniformly spaced grid in $\log_{10}(\lambda)$, we shifted each spectrum to the stellar rest frame using the relativistic Doppler relation,
\[
1+z \;=\; \sqrt{\frac{1+\beta}{1-\beta}}, \qquad \beta=\frac{v}{c},
\]
where $v$ is the heliocentric radial velocity (RV) from the \texttt{RV\_ADOP} column and $c$ is the speed of light in vacuum. As is standard in SDSS data processing \citep{Bolton2012}, the transformation from the observed frame to the rest frame in logarithmic wavelength coordinates reduces to a simple additive shift:
\[
\log_{10}\lambda_{\mathrm{rest}} \;=\; \log_{10}\lambda_{\mathrm{obs}} \;-\; \log_{10}(1+z).
\]
Operationally, for each spectrum $i$ we compute the scalar shift $\Delta_i=\log_{10}(1+z_i)$. The $\log_{10}\lambda$ axis is then shifted by $-\Delta_i$ and the flux is interpolated back onto the common grid using cubic interpolation.

\begin{figure}[ht]
    \centering
    \includegraphics[width=0.40\textwidth]{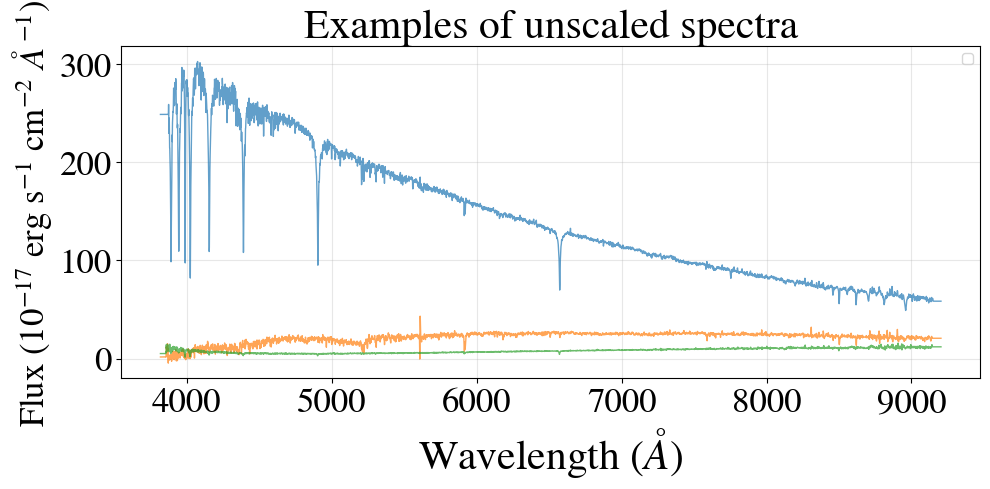} % Ajustei para 0.8 para melhor visibilidade
    \caption{Examples of unscaled SDSS spectra illustrating the large variation in absolute flux levels, which range from $\sim 10$ to $300 \times 10^{-17}$ erg s$^{-1}$ cm$^{-2}$ \AA$^{-1}$. This is primarily driven by extrinsic factors, such as source distance, rather than intrinsic atmospheric properties.}
    \label{fig:unscaled_spectra}
\end{figure}

This correction ensures that all spectral absorption features are aligned at their respective wavelengths, allowing the model to learn physically meaningful local spectral features. However, as shown in Figure \ref{fig:unscaled_spectra}, the absolute flux levels vary substantially across the sample. These variations are driven largely by extrinsic factors, particularly source distance, rather than by target atmospheric parameters themselves. To isolate the relevant physical information and to prevent the model from relying on absolute flux intensity, a per-spectrum normalization step was therefore applied.

To this end, each spectrum is standardized \emph{individually} by subtracting its mean value and scaling by its own standard deviation. This normalization allows the model to focus on the relative strengths of absorption lines and the overall spectral shape rather than absolute flux intensities. For a spectrum $\mathbf{x}_i \in \mathbb{R}^{p}$ (row $i$ of the input matrix $X$), the transformation is defined as:
\[
\begin{gathered}
\bar{x}_i = \frac{1}{p}\sum_{j=1}^{p} x_{ij},
\qquad
s_i = \sqrt{\frac{1}{p}\sum_{j=1}^{p}\bigl(x_{ij}-\bar{x}_i\bigr)^2}, \\[4pt]
\tilde{x}_{ij} = \frac{x_{ij}-\bar{x}_i}{s_i},
\end{gathered}
\]
where $p = 4000$ represents the number of spectral pixels.

\begin{figure}[ht]
    \centering
    \includegraphics[width=0.30\textwidth]{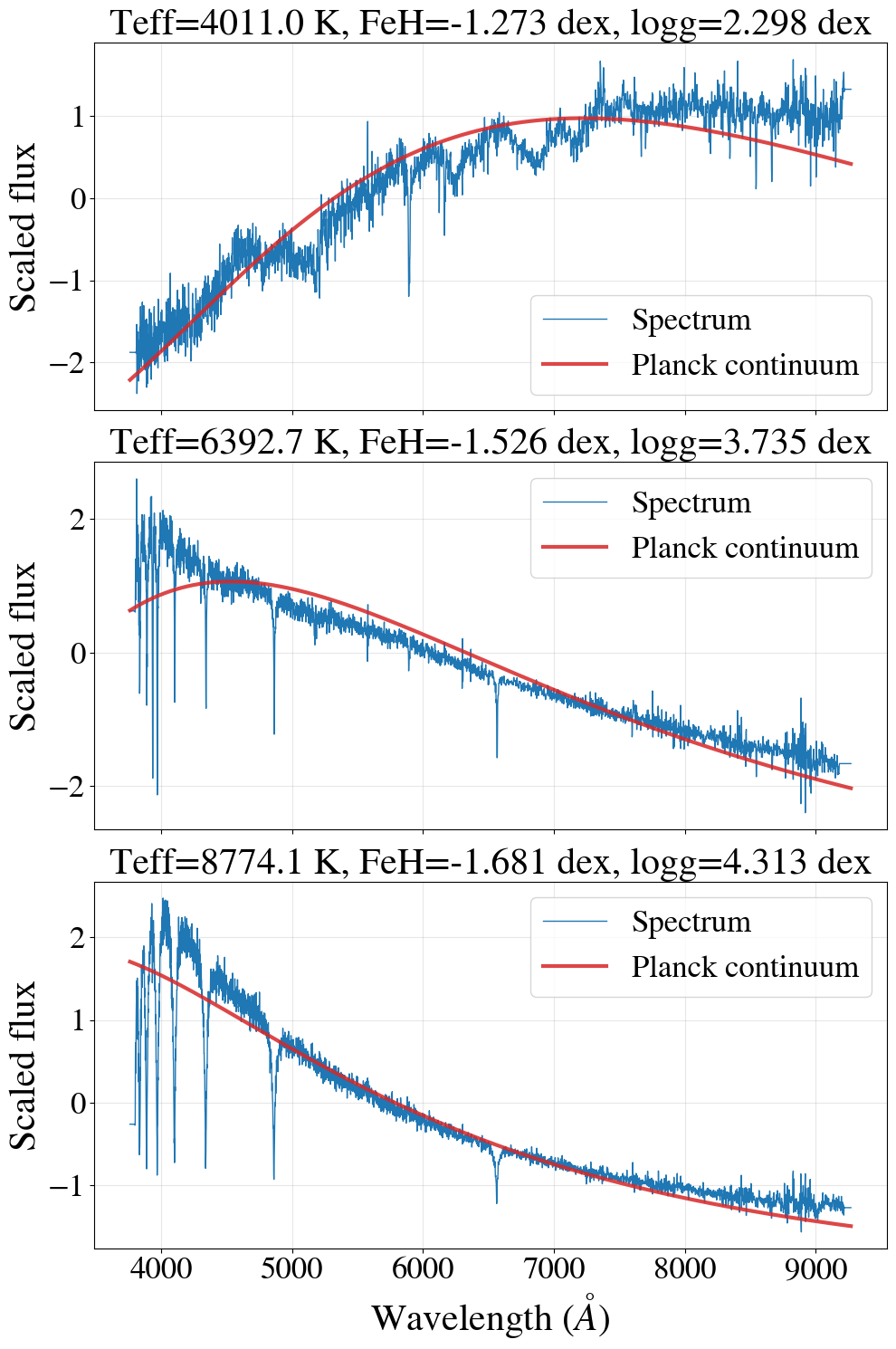}
    \caption{Examples of normalized SDSS spectra after wavelength alignment and per-spectrum standardization. The red curves denote the theoretical Planck continua for the corresponding $T_{\mathrm{eff}}$ values, illustrating that the normalization preserves the overall spectral shape.}
    \label{fig:normalized_spectra}
\end{figure}

The same transformation is applied to each spectrum in all data partitions. As illustrated in Figure \ref{fig:normalized_spectra}, the normalization largely removes scale differences associated with distance, bringing all spectra to a common dimensionless scale centered at zero.

The three target variables have different scales and marginal distributions, so each one is scaled independently using \texttt{RobustScaler} from scikit-learn \citep{Pedregosa2011}, which centers the data by the median and scales it by the interquartile range (IQR). The scaling parameters estimated from the training set are then applied to the validation and test sets.

To improve model robustness and mitigate overfitting, we employed data augmentation to expand the training set. New samples were generated by adding Gaussian noise to each spectrum in order to simulate detector and photon-noise effects, that is,
\[
\epsilon \sim \mathcal{N}\!\left(0,(l_{\mathrm{noise}}\sigma_S)^2\right), 
\qquad
S' = S + \epsilon,
\]
where $S$ denotes the original spectrum, $\sigma_S$ its standard deviation, and $l_{\mathrm{noise}}$ a noise-level factor.

\section{Model Definition}
\label{sec:model}

We consider the problem of learning a nonlinear mapping from an input spectrum $\mathbf{x}\in\mathbb{R}^{p}$ (flux on a fixed $\log\lambda$ grid) to the three target atmospheric parameters
$\mathbf{y}=\big(T_{\mathrm{eff}},\,[\mathrm{Fe/H}],\,\log g\big)^\top\!\in\mathbb{R}^{3}$\@. To this end, we use a compact residual multitask MLP composed of a shared residual backbone followed by three task-specific heads \citep{He2016_ResNet}.

\begin{figure}[ht]
    \centering
    \includegraphics[width=0.33\textwidth]{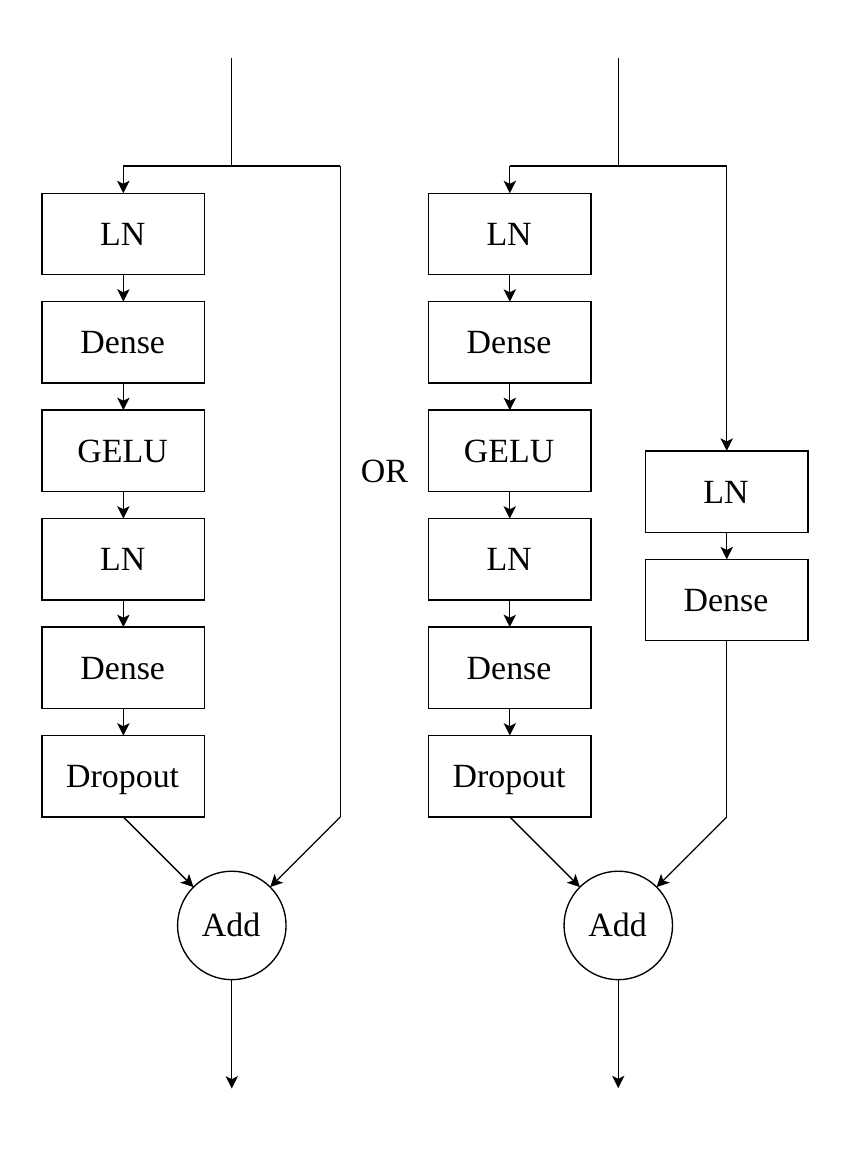}
    \caption{Schematic representation of the residual block used in the shared backbone. The main branch consists of two dense layers with Layer Normalization, GELU activation, and dropout. The shortcut path is either the identity when $d_{b-1}=d_b$ or a Layer Normalization step followed by a linear projection when $d_{b-1}\neq d_b$. The outputs of the two branches are then summed to produce $\mathbf{h}^{(b)}$.}
    \label{fig:residual_block}
\end{figure}

The input is mapped to an initial hidden representation as follows:
\[
\mathbf{h}^{(0)} = \phi\!\left(\mathrm{LN}(\mathbf{X}W_0 + \mathbf{b}_0)\right),
\]
where $\phi(\cdot)$ denotes a pointwise activation function and $\mathrm{LN}(\cdot)$ denotes the Layer Normalization operation \citep{ba2016layernormalization}. The shared backbone consists of $B$ residual blocks defined as:
\[
\begin{aligned}
\mathbf{u}^{(b)} &= \mathrm{LN}\!\left(\mathbf{h}^{(b-1)}\right)W_1^{(b)} + \mathbf{b}_1^{(b)},\\[4pt]
\mathbf{v}^{(b)} &= \mathrm{LN}\!\left(\phi\!\left(\mathbf{u}^{(b)}\right)\right),\\[4pt]
\mathbf{t}^{(b)} &= \mathrm{Dropout}\!\left(\mathbf{v}^{(b)}W_2^{(b)} + \mathbf{b}_2^{(b)}\right),\\[4pt]
\mathbf{r}^{(b)} &=
\begin{cases}
\mathbf{h}^{(b-1)}, & \text{if } d_{b-1}=d_b,\\[4pt]
\mathrm{LN}\!\left(\mathbf{h}^{(b-1)}\right)W_r^{(b)} + \mathbf{b}_r^{(b)}, & \text{if } d_{b-1}\neq d_b,
\end{cases}\\[6pt]
\mathbf{h}^{(b)} &= \mathbf{t}^{(b)}+\mathbf{r}^{(b)}.
\end{aligned}
\]
for $b=1,\dots,B$, where $d_{b-1}$ and $d_b$ denote the input and output dimensions of block $b$. When the dimensionality changes, the shortcut is matched by a Layer Normalization step followed by a learned linear projection, as illustrated in Figure~\ref{fig:residual_block}.

After the $B$ residual blocks, the shared representation is further refined by an additional LN operation,
\[
\mathbf{s} = \mathrm{LN}\big(\mathbf{h}^{(B)}\big),
\]

Each task-specific head processes the shared representation $s$ through an MLP. Let $\mathbf{q}_i^{(0)}={\mathbf{s}}$, and define
\[
\mathbf{q}_i^{(\ell)}=
\phi\!\left(\mathbf{q}_i^{(\ell-1)}A_i^{(\ell)}+\mathbf{a}_i^{(\ell)}\right),
\qquad \ell=1,\dots,L_i,
\]
where $A_i^{(\ell)}$ and $\mathbf{a}_i^{(\ell)}$ are the weight matrix and bias vector of the $\ell$-th hidden layer of the $i$-th task head, $\phi(\cdot)$ is its activation function, and dropout may be applied after each hidden transformation. The final scalar prediction for task $i$ is given by
\[
\hat{y}_i = \mathbf{q}_i^{(L_i)}\mathbf{w}_i^{(o)} + b_i^{(o)},
\]
where $\mathbf{w}_i^{(o)}$ and $b_i^{(o)}$ are the parameters of the output layer. In this work, the three task-specific outputs correspond to the stellar parameters.

We chose the Gaussian Error Linear Unit (GELU) as the activation function $\phi$, since it provides a smooth nonlinear transformation and tends to preserve small but informative inputs rather than completely suppressing them \citep{Hendrycks2016}, which can be advantageous when modeling subtle variations in stellar spectra.

\section{Training}
\label{sec:train}

To determine a suitable architecture and hyperparameters, we used the Keras Tuner library \citep{OMalley2019_KerasTuner} within the Keras/TensorFlow framework \citep{Chollet2015,Abadi2016}. We defined a flexible search space over the shared backbone and task-specific heads and employed Bayesian optimization \citep{Snoek2012} over 100 trials to identify the configuration that minimized the validation loss.

The hyperparameter search space was defined as follows: the width of the initial stem dense layer ranged from 64 to 128 units; the shared residual backbone ranged from 1 to 2 residual blocks, with 32 to 64 units per block; and the task-specific heads were selected from predefined topology templates ranging from a single 16-unit layer to a deeper 48-32-16 configuration. To mitigate overfitting, dropout rates were tuned independently for the stem ($0.0 - 0.5$), trunk ($0.0 - 0.4$), and task-specific heads ($0.0 - 0.4$). 

To handle the multi-variable nature of the estimation problem, the training objective was defined as a weighted sum of task-specific Huber losses ($\delta=1.0$) \citep{10.1214/aoms/1177703732}. The Huber loss was chosen because it behaves quadratically for small residuals and linearly for large residuals, thereby combining sensitivity near the optimum with reduced sensitivity to outliers. The contribution of each task was weighted by the inverse of its empirical variance in the training set, in order to balance the optimization across the three target parameters.

Model parameters were optimized using the AdamW algorithm, which incorporates decoupled weight decay regularization \citep{loshchilov2019decoupledweightdecayregularization}. The learning rate and weight decay were treated as hyperparameters and sampled on a logarithmic scale from $[10^{-4}, 5 \times 10^{-3}]$ and $[10^{-5}, 10^{-2}]$, respectively.

For each original spectrum in the training set, two additional augmented samples were generated by adding Gaussian noise with noise levels drawn uniformly between 1\% and 5\%. This resulted in an augmented training set containing $3 \times 30{,}000 = 90{,}000$ spectra.

During training, we used a learning-rate schedule consisting of a linear warm-up phase for the first 5\% of total training steps, followed by a monotonic cosine decay \citep{loshchilov2017sgdrstochasticgradientdescent} down to a minimum threshold. The batch size and number of epochs were set to 256 and 30, respectively. The best-performing model weights were retained via checkpointing.

\section{Results and Evaluation}
\label{sec:Results}

The Bayesian optimization procedure selected an architecture containing 542,771 trainable parameters. The best-performing configuration consisted of an initial stem layer with 128 units and no dropout, followed by a shared trunk with one residual block of 64 units and a dropout rate of 0.4.

The selected task-specific heads had distinct topologies, reflecting the underlying different ways in which each parameter affects the spectra. The effective temperature ($T_{\mathrm{eff}}$) head required the deepest configuration, with 48- and 32-unit layers, whereas the surface gravity ($\log g$) used a single 48-unit layer. The metallicity ($[\mathrm{Fe/H}]$) head required only a 16-unit linear projection. Dropout in the task-specific heads was selected as zero, indicating that the stronger regularization applied in the shared trunk, together with a decoupled weight decay of 0.01, was sufficient. The optimal learning rate was found to be $7.0 \times 10^{-4}$.

\begin{figure}[ht]
    \centering
    \includegraphics[width=0.27\textwidth]{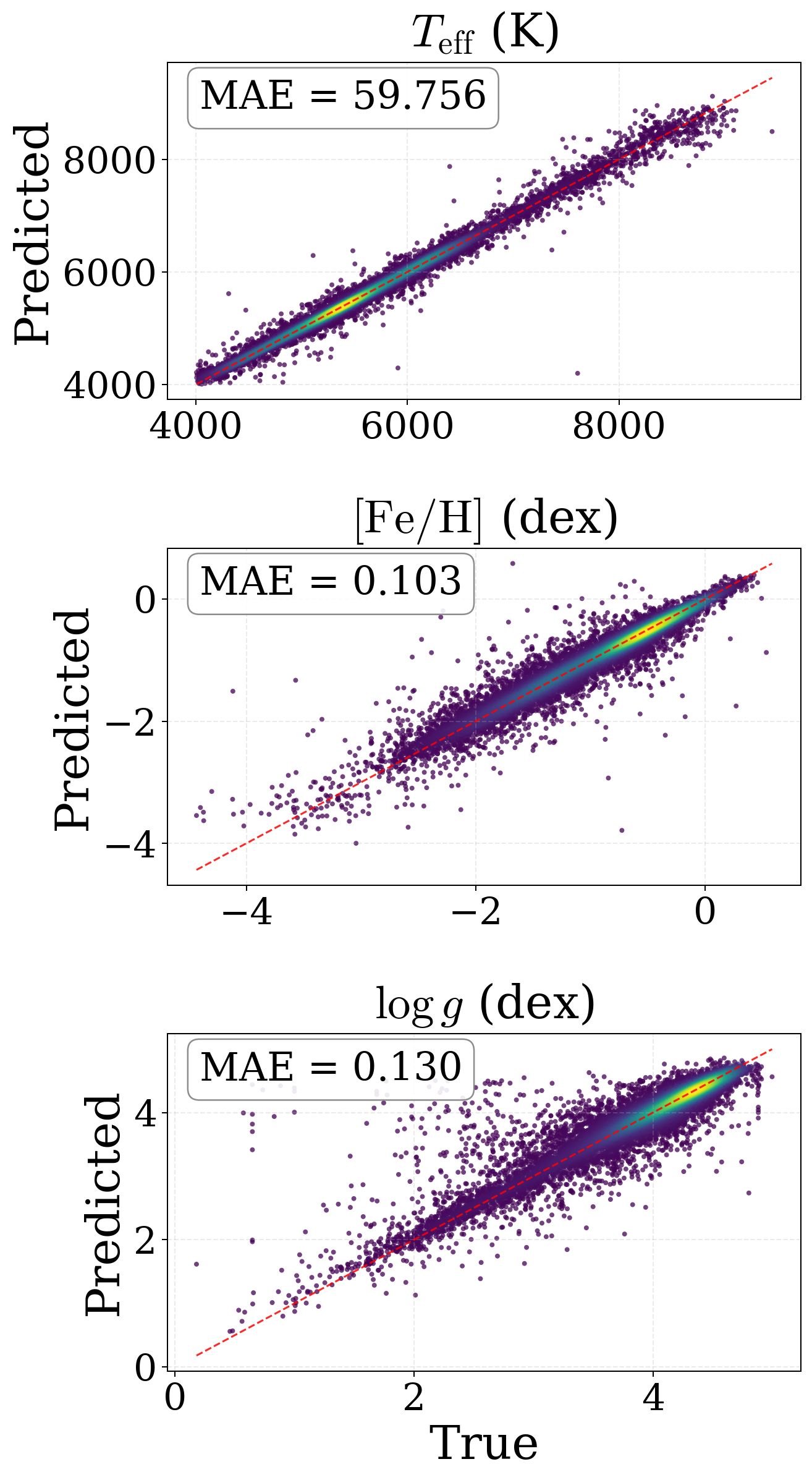}
    \caption{Predicted versus true values for $T_{\text{eff}}$, $[\text{Fe}/\text{H}]$, and $\log g$ on the test set, with point density represented using kernel density estimation.
The dashed red line indicates perfect prediction.}
    \label{fig:scatter}
\end{figure}

On the test set, the proposed model showed good predictive performance, with predictions concentrated around the identity line in Figure \ref{fig:scatter}. When normalized against the empirical range of each parameter in the test set (approximately 5441~K for $T_{\mathrm{eff}}$, 4.97~dex for $[\mathrm{Fe/H}]$, and 4.81~dex for $\log g$), the corresponding errors were 1.10\%, 2.07\%, and 2.70\%, respectively. 

To compare the proposed approach with simpler alternatives, we benchmarked it against standard linear estimators: Ordinary Least Squares (OLS) \citep{hastie_09_elements-of.statistical-learning} and Ridge Regression \citep{Hoerl01021970}. As shown in Table \ref{tab:baselines}, the linear models struggle to capture the complex relationship between spectral features and the target parameters. The proposed network reduces these errors by more than 50\% across all three physical parameters, suggesting that the mapping from spectral flux to stellar atmospheric parameters is strongly non-linear.

\begin{table}[htpb]
\centering
\caption{Performance comparison between the linear baselines and the proposed model on the test set.}
\label{tab:baselines}
\resizebox{\linewidth}{!}{%
\begin{tabular}{lccc}
\toprule
\textbf{Method} & \textbf{MAE $T_{\mathrm{eff}}$ (K)} & \textbf{MAE $[\mathrm{Fe/H}]$ (dex)} & \textbf{MAE $\log g$ (dex)} \\
\midrule
Ordinary Least Squares & 143.75 & 0.26 & 0.28 \\
Ridge Regression & 115.24 & 0.21 & 0.23 \\
\textbf{Proposed NN} & \textbf{59.75} & \textbf{0.10} & \textbf{0.13} \\
\bottomrule
\end{tabular}%
}
\end{table}

For a fair comparison with alternative neural architectures, we evaluated the proposed model against two additional baselines trained under the same experimental setting: (i) a CNN architecture based on \citep{10.1093/mnras/stx3298} and (ii) a Deep Neural Network (DNN) following \citep{2017RAA1736L}. All models were trained and evaluated using identical data splits and preprocessing procedures. The results are summarized in  Table \ref{tab:sota_comparison}. Notably, the DNN requires approximately 13.7 million parameters, since it models each stellar parameter with an independent network, in contrast to the proposed multitask formulation.

\begin{table}[htpb]
\centering
\caption{Test-set performance and model complexity comparison across alternative neural network architectures.}
\label{tab:sota_comparison}
\resizebox{\columnwidth}{!}{%
\begin{tabular}{lcccc}
\toprule
\textbf{Model} & \textbf{Params} & \textbf{MAE $T_{\mathrm{eff}}$ (K)} & \textbf{MAE $[\mathrm{Fe/H}]$ (dex)} & \textbf{MAE $\log g$ (dex)} \\
\midrule
CNN & 4.13M & 81.72 & 0.138 & 0.166 \\
DNN & 13.66M & 60.51 & \textbf{0.103} & 0.140 \\
\textbf{Proposed model} & \textbf{0.55M} & \textbf{59.75} & 0.103 & \textbf{0.130} \\
\bottomrule
\end{tabular}%
}
\end{table}

For a more direct comparison, Table~\ref{tab:relative_comparison} reports the relative values of the evaluated metrics, normalized with respect to the DNN model (= 1.00). The DNN achieves a slightly lower MAE for $T_{\mathrm{eff}}$; however, the proposed architecture attains comparable performance across all three parameters with substantially lower model complexity and inference time. In contrast, the CNN model exhibits higher errors for all three atmospheric parameters.
\begin{table}[htpb]
\centering
\caption{Relative performance, complexity, and average inference time normalized with respect to the DNN baseline. Time corresponds to the average inference time measured on the full test set (15,000 spectra), averaged over 10 runs. Values below 1 indicate improvement over the DNN.}
\label{tab:relative_comparison}
\resizebox{\columnwidth}{!}{%
\begin{tabular}{lccccc}
\toprule
\textbf{Model} & \textbf{Params} & \textbf{Time} & \textbf{MAE $T_{\mathrm{eff}}$} & \textbf{MAE $[\mathrm{Fe/H}]$} & \textbf{MAE $\log g$} \\
\midrule
CNN & 0.30 & 0.50 & 1.35 & 1.34 & 1.18 \\
DNN  & 1.00 & 1.00 & 1.00 & 1.00 & 1.00 \\
\textbf{Proposed NN} & 0.04 & 0.16 & 0.99 & 1.00 & 0.93 \\
\bottomrule
\end{tabular}%
}
\end{table}

The ablation results in Table \ref{tab:relative_comparison_ablation} clarify the contribution of the main architectural components. Removing the skip connections leads to a clear degradation in performance across all targets, showing that they play an important role in learning effective representations. When LN is also removed, the degradation becomes even more pronounced, particularly for $T_{\mathrm{eff}}$ and $\log g$, indicating that normalization improves training stability. Single-task models achieve similar accuracy for individual parameters, but require nearly three times more parameters and longer inference time, indicating that the multitask formulation provides a more efficient overall solution.

\begin{table}[htpb]
\centering
\caption{Relative performance, model complexity and average inference time normalized with respect to the proposed model. Values below 1 indicate improvement over this baseline.}
\label{tab:relative_comparison_ablation}
\resizebox{\columnwidth}{!}{%
\begin{tabular}{lccccc}
\toprule
\textbf{Model} & \textbf{Params} & \textbf{Time} & \textbf{MAE $T_{\mathrm{eff}}$} & \textbf{MAE $[\mathrm{Fe/H}]$} & \textbf{MAE $\log g$} \\
\midrule
No Skip & 0.99 & 0.69 & 1.39 & 1.24 & 1.15 \\
No Skip + No LN  & 0.98 & 0.73& 1.56 & 1.17 & 1.28 \\
No Multitasking & 2.97 & 2.71 & 0.99 & 1.00 & 1.01 \\
\textbf{Proposed NN} & 1.00 & 1.00 & 1.00 & 1.00 & 1.00 \\
\bottomrule
\end{tabular}%
}
\end{table}

When both skip connections and LN are removed, the resulting architecture reduces to a standard feedforward network, resembling the DNN formulation proposed by \cite{2017RAA1736L}. However, because our setting uses substantially fewer units per layer, this simplified variant yields markedly worse performance.

\section{conclusion}
\label{sec:Conclusion}

This work shows that accurate stellar parameter estimation from SDSS spectra can be achieved with compact neural architectures. The results indicate that the proposed residual multitask MLP attains performance comparable to deeper neural baselines while requiring substantially fewer parameters and lower inference time. These findings suggest that compact architectures can effectively capture the relevant nonlinear relationships between spectral features and atmospheric parameters. This is particularly relevant for large-scale spectroscopic pipelines, where computational efficiency and scalability are essential.

\section*{Acknowledgments}
The authors thank Prof. Laerte Sodré Júnior for the valuable discussions and suggestions throughout the development of this work.

\bibliography{ifacconf}

\end{document}